\documentclass[%
 amsmath,amssymb,
 reprint,%
]{revtex4-1}

\usepackage{graphicx}
\usepackage{dcolumn}
\usepackage{bm}

\usepackage[utf8]{inputenc}
\usepackage[T1]{fontenc}
\usepackage{mathptmx}
\usepackage[colorlinks]{hyperref}

\begin{document}

\preprint{AIP/123-QED}

\title[]{Visible and Near-infrared Microdisk Resonators on a 4H-Silicon-Carbide-on-Insulator Platform}

\author{Chengli Wang}
\author{Chen Shen}
\author{Ailun Yi}
\author{Liping Zhou}
\author{Yifan Zhu}
\author{Kai Huang}
\author{Sannian Song}
\author{Min Zhou}
\author{Jiaxiang Zhang}
\email{Jiaxiang.zhang@mail.sim.ac.cn}
\author{Xin Ou}
\email{xin.ou@mail.sim.ac.cn}
\affiliation{ 
State Key Laboratory of Functional Materials for Informatics, Shanghai Institute of Microsystem and Information Technology, Chinese Academy of Sciences, Shanghai, 200050, China
}%
\author{Shumin Yang}
\affiliation{%
Shanghai Synchrotron Radiation Facility, Shanghai Advanced Research Institute, Chinese Academy of Sciences, Shanghai 201204, China
}%

\date{\today}

\begin{abstract}
Wavelength-sized microdisk resonators were fabricated on a single crystalline 4H-silicon-carbide-on-insulator platform (4H-SiCOI). By carrying out micro-photoluminescence measurements at room temperature, we show that the microdisk resonators support whispering-gallery modes (WGMs) with quality factors up to 5.25$\times$10$^3$ and mode volumes down to 2.61$\times(\lambda/n)^3$ at the visible and near-infrared wavelengths. Moreover, the demonstrated wavelength-sized microdisk resonators exhibit WGMs whose resonant wavelengths compatible with the zero-phonon lines of silicon related spin defects in 4H-SiCOI, making them a promising candidate for applications in cavity quantum electrodynamics and integrated quantum photonic circuits.
\end{abstract}
\maketitle
\section{\label{sec:level1} Introduction}
Silicon carbide (SiC) has recently emerged as a novel material choice for applications in quantum and nonlinear photonics\cite{cardenas2015optical,zheng20194h,guidry2020optical,song2019ultrahigh,son2020developing}. It offers many outstanding material properties highly complementary to conventional silicon-based photonics, including large transmission window from the ultraviolet to far infrared, high refractive index , large second- and third-order optical nonlinearities, superior thermal and chemical stability\cite{singh1971nonlinear,harris1995properties}. In particular, SiC harbors a rich assortment of optically addressable solid-state spin defects\cite{castelletto2020silicon}. By combining the abundant spin defects with nanophotonic devices, SiC is able to serve as a single material platform to develop linear and nonlinear optical systems, novel quantum devices and especially integrated quantum photonic circuits (IQPC).\\
\indent~In analogous to diamond, realization of optical microcavities on SiC material platform are particularly appealing, and they facilitate the exploration of promising defect-cavity systems where the light extraction efficiency of the zero-phonon line (ZPL) emission, as well as the efficiency of the coherent spin state readout, can be significantly enhanced via the Purcell effect\cite{purcell1995spontaneous}.Currently, numerous polytypes of SiC are under investigation. Cubic SiC (3C-SiC) and hexagonal 4H-SiC are among the most suitable material platforms owning to their well-established techniques for wafer-scale sample growth and processing. Since 3C-SiC can be directly grown on a silicon substrate though epitaxial growth, this material enables demonstration of a variety of optical microcavities\cite{cardenas2013high,fan2020high,lu2020silicon,allioux2018toward}. Nevertheless, the lattice mismatch between Si and 3C-SiC could induce a high density of growth defects at the Si-SiC interface, making 3C-SiC platform impractical for single spin-defects based quantum applications \cite{calusine2016cavity}. In contrast, 4H-SiC offers an alternative for defect-based quantum applications. Its great advantage manifests as the feasibility of integrating various silicon-related spin defects in the visible and near infra-red range\cite{castelletto2020silicon}. Until now, tremendous efforts have been devoted to realize spin-cavity system in bulk 4H-SiC material\cite{crook2020purcell}. Despite the successful demonstration, the reported cavity geometry in bulk 4H-SiC strongly relies on sophisticated carving techniques.
To circumvent the complication of using carving techniques in bulk material, a long-proposed solution is to use 4H-SiC thin-film-on-insulator (4H-SiCOI). The recent work of realizing high-quality 1D PhC nanobeam microcavity coupled with a single silicon vacancy on 4H-SiCOI\cite{lukin20204h} represents an important progress in this direction. However, the work used 4H-SiC slabs that were prepared by thinning bulk wafers. Such process requires substrate grinding through mechanical thinning and chemical polishing. \\
\indent~Compared with the bulk wafer thinning technique, applying the ion cut process to prepare 4H-SiCOI is practically favorable. Considerable benefits include the mass production of wafer-scale 4H-SiCOI with well-controlled thickness and the industry-compatible device fabrication method\cite{yi2020wafer}. With these manifold advantages, there has been an increasing interest on developing nanophotonic cavities in 4H-SiCOI by the ion cut process in recent years\cite{zheng20194h,zheng2019high}. However, all the reported devices were operated at telecom wavelengths. The existing big discrepancy between the resonance wavelengths and the spin-defects in 4H-SiC impedes an effective coupling of the cavity resonance and the defects in those devices. Towards the realization of a defect-cavity system, the development of microcavity structure that is capable of working at the ZPL emission wavelengths of color centers in 4H-SiCOI is highly desirable.\\
\indent~Here we present the fabrication of wavelength-scale microdisk resonators on a 4H-SiCOI material platform formed by the ion cut process. By carrying out micro-photoluminescence ($\mu$-PL) experiment, we characterized the quality factors up to  5.25$\times$10$^3$ and the simulated mode volume of 2.61$\times$($\lambda$/n)$^3$ in visible to near-infrared range, which shows the potential for realizing efficient defect-cavity system in 4H-SiCOI.
\section{\label{sec:level2} Experiment and discussions}
\begin{figure}[hbtp]
\centering
\fbox{\includegraphics{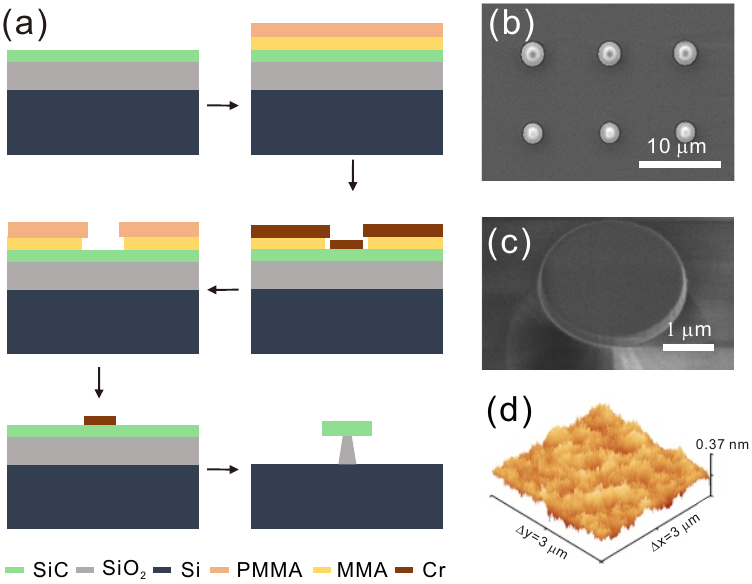}}
\caption{(a) Fabrication process flow of the wavelength-sized microdisk resonators in 4H-SiCOI. (b) SEM image of the fabricated microdisk array with different radius. (c) A zoomed-in SEM image of a single microdisk resonator featuring with a radius of 1.5 $\mu$m and thickness of 160 nm.(d) 3D atomic force microscopy scan of the top surface of SiCOI.}
\label{fig1}
\end{figure}

\begin{figure*}[htbp]
\centering
\fbox{\includegraphics[scale =1.0]{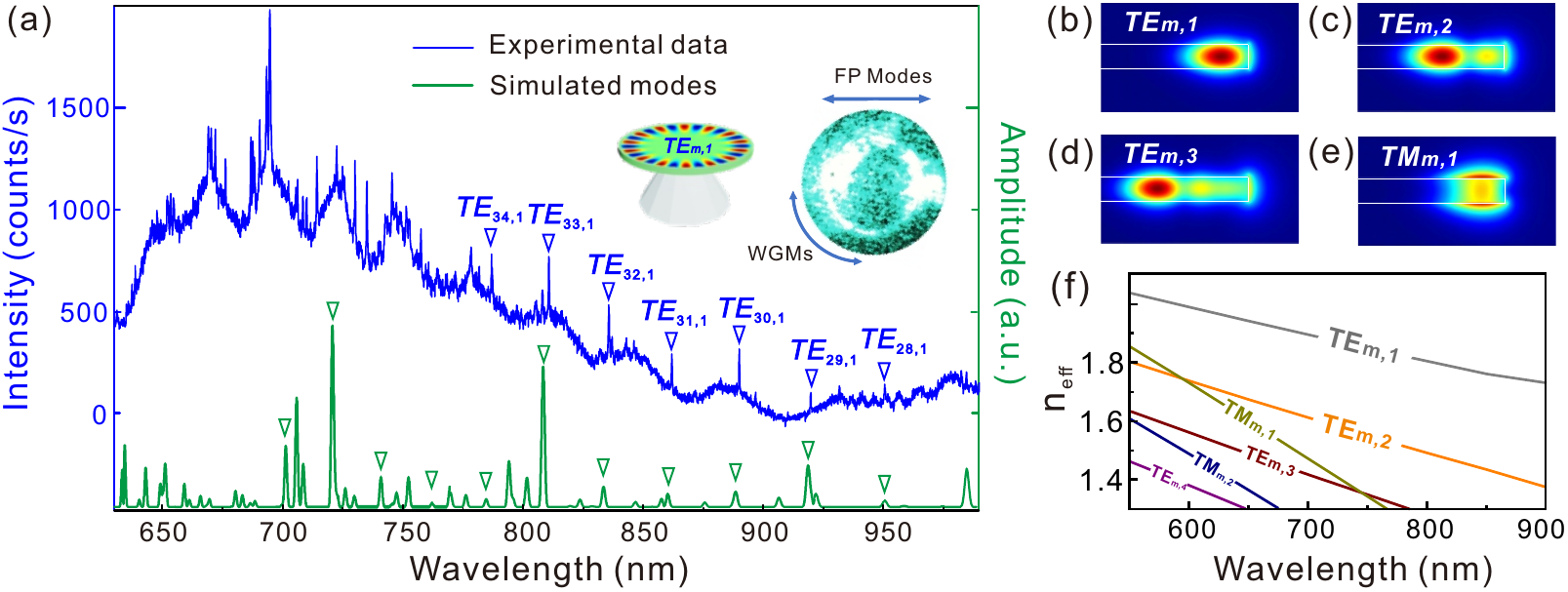}}
\caption{(a) Experimentally measured PL spectrum (blue line) and simulated resonant spectrum (green line) of the 2 $\mu$m-radius microdisk resonator. Wavelength-dependent electric-field profiles of $TE_{m,1}$, $TE_{m,2}$, $TM_{m,1}$, and $TE_{m,3}$ in the 2 $\mu$m-radius microdisk resonator, which located at (b) 862.1 nm, (c) 740 nm, (d) 700 nm and (e) 650 nm, respectively. (f) Simulated effective index of the $TE_{m,1}$, $TE_{m,2}$, $TM_{m,1}$, $TE_{m,3}$, and $TM_{m,2}$ at different wavelengths in the 2 $\mu$m-radius microdisk resonator.}
\label{fig2}
\end{figure*}

Fabrication process of the microdisk resonators in 4H-SiCOI is schematically shown in Figure \ref{fig1}(a). Firstly, a heterogeneous 4H-SiCOI wafer was prepared by the ion cut process, and the detailed fabrication process of this heterogeneous wafer was presented elsewhere\cite{yi2020wafer}. The wafer consists of a 4H-SiC layer with a thickness of 160 nm, a buried silicon dioxide layer with a thickness of 2 $\mu$m and a silicon supporting substrate. To fabricate the microdisk, a 300 nm thick methyl methacrylate (MMA) layer and a 300 nm thick poly methyl methacrylate (PMMA) layer were consecutively spin-coated onto the 4H-SiCOI substrate. We then defined the microdisk structures by using an electron-beam lithography tool (EBPG 5200). After developing in methyl isobutyl ketone (MIBK): isopropyl alcohol (IPA) in a ratio of 3:1 for 45s, a 100 nm thick Cr layer serving as a hard mask was deposited. In the next step, we employed an Inductively Coupled Plasma-Reactive Ion Etching (ICP-RIE) tool to dry-etch the microdisk structures. The residual Cr mask was removed by chromium etchant. Finally, the buried oxide layer was precisely etched by diluted hydrofluoric (HF) acid to form a supporting pedestal. Figure \ref{fig1}(b) shows a scanning electron microscopy (SEM) image of the fabricated 4H-SiC microdisk resonators a radius varying from 1 to 2 $\mu$m and Figure \ref{fig1}(c) shows a zoomed-in SEM image of a microdisk with a fixed radius of 1.5 $\mu$m. Noteworthy to mention is the surface roughness of the microdisk resonator as shown in Figure \ref{fig1}(d). Before fabricating the device, a refined chemical-mechanical polishing was employed to reduce the surface roughness down to 0.3 nm. Such high surface quality has proven to be better than the surface treated by the wet oxidation process used in previous work\cite{zheng20194h}.
\\
\indent~Experimentally, we characterize the optical properties of these wavelength-sized microdisk resonators by employing a room temperature $\mu$-PL setup (for details, see Supplement 1). Optical excitation was performed by using a continuous wave 532 nm laser which was focused atop of microdisk through a near-infrared-infinity corrected objective (Mitutoyo NIR 50$\times$). PL signal was then collected by the same objective and delivered to a high-resolution spectrometer (SP2750) equipped with a thermoelectric-cooled charge-coupled device. Figure \ref{fig2}(a) shows the PL spectrum (blue line) from a microdisk with a radius of 2 $\mu$m.  The broad background emission is most likely from the extrinsic oxygen-related defects near the SiC surface and the phonon replicas of silicon vacancy centers. Intensive studies have revealed that, in proton-irradiated 4H-SiCOI material, the oxygen-related defects dominate background emission in the visible range between 600 nm and 750 nm, and the phonon replicas dominate the near-infrared range \cite{fuchs2013silicon,ruhl2018controlled,lohrmann2015single}.\\
\indent~From the spectrum shown in Figure \ref{fig2}(a), we observe a set of sharp peaks in a wide range from 650 nm to 1000 nm, and their origin can be ascribed to the whispering gallery mode (WGMs) excited in the microdisk. In order to identify them, we carried out an optical simulation by means of a three-dimensional FDTD method. The refractive index of 4H-SiC is set according to the previous report\cite{wang20134h}, and the simulated radius is same as the fabricated microdisk. The simulated spectrum is shown in Figure \ref{fig2}(a) (green line). In the near infrared spectral range (775-1000 nm), a remarkably good agreement between the experimental results and the simulated resonant frequencies is observed. Further analysis on the electric field profiles shown in Figure \ref{fig2}(b) suggests that, in the  near infrared range, the WGMs are dominated by the transverse electric modes ($TE_{m,n}$) with the radial mode n = 1 and the azimuthal mode numbers m = 28\textasciitilde34. As compared to the $TE_{m,1}$, transverse magnetic modes (TM) or higher-order radial TE only appear in the visible range and they contribute more complex spectral lines. To interpret the above phenomenon, we performed a mode analysis and calculated the effective refractive index ($n_{eff}$) for WGMs in the microdisk. The mode species of the first mode families are identified and labeled as $TE_{m,1}$, $TE_{m,2}$, $TM_{m,1}$, $TE_{m,3}$, $TM_{m,2}$ and $TE_{m,4}$ in Figure \ref{fig2}(f). It is shown that as the wavelength decreases, $n_{eff}$ for the WGMs with higher order radial number increases dramatically.  Especially for the visible wavelengths from 650 nm to 775 nm, the $n_{eff}$ of $TE_{m,2}$, $TM_{m,1}$, and $TE_{m,3}$ increases to above 1.5, suggesting that these modes can acquire sufficient modal confinement and become resonant modes in the visible region. This can explain the phenomenon observed in the experimental and simulated spectrum, that is, relatively clean $TE_{m,1}$ modes appear in the near-infrared region and higher-order radial TE and whisper gallery TM modes such as $TE_{m,2}$, $TM_{m,1}$, and $TE_{m,3}$ occur in the visible region. Figures \ref{fig2}(c-e) illustrate the electric field profile of the  $TE_{m,2}$, $TM_{m,1}$, and $TE_{m,3}$ modes. 
\indent~In addition to the sharp resonant peaks, the spectrum shows a number of broad oscillating peaks. To find their derivations, we took a microscopic image of the optically excited microdisk as shown in the inset of Figure \ref{fig2}(a). We found two bright spots located on the side wall of the microdisk. This  is due to the resonance phenomenon caused by the Fabry-Perot (FP) cavity formed between the two sidewalls of the microdisk. A similar phenomenon is also observed in a wavelength-size GaP microdisk resonator\cite{barclay2009chip}. The free spectrum range (FSR) of the FP resonance is evaluated by $FSR = \lambda^2/2dn = 30.2$ nm, with $\lambda = 800$ nm being the wavelength, \emph{d} = 4 $\mu$m the diameter of the microdisk and $n = 2.65$ the refractive index of the 4H-SiC\cite{wang20134h}.
\begin{figure}[htbp]
\centering
\fbox{\includegraphics[scale = 1.0]{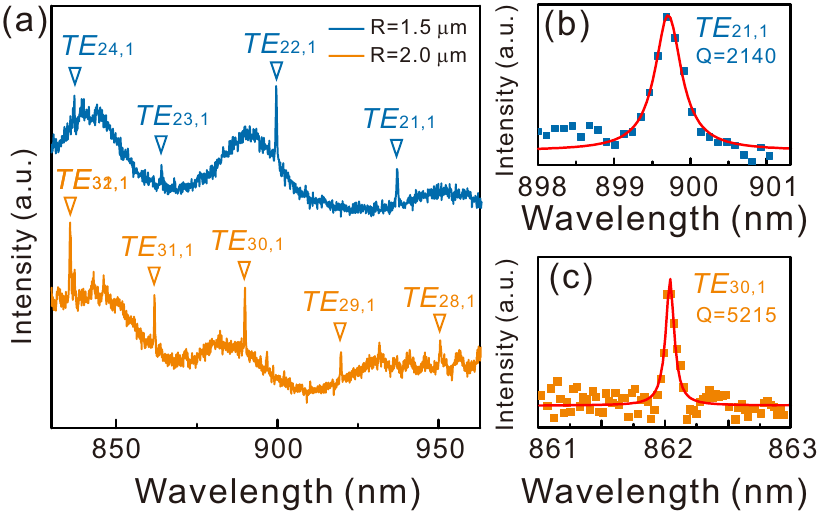}}
\caption{(a) Measured PL spectrum from microdisk resonators with a radius of 1.5 $\mu$m (dark blue) and 2 $\mu$m (yellow) respectively. WGMs are denoted by the triangles. Normalized PL spectra and theoretic fits of the cavity modes located at 899.7 nm (b) in the 1.5 $\mu$m-radius microdisk resonator and at 862.1 nm (c) in the 2 $\mu$m-radius microdisk resonator.}
\label{fig3}
\end{figure}
\indent~Figure \ref{fig3}(a) shows the PL spectrum of two microdisks with a radius of 1.5 and 2 $\mu$m respectively. Both of them show the same characteristics except for the different FSR. By taking the geometric structure of these two microdisks into account, we find the FSR of them are about 30 nm and 40 nm. Moreover, we select the optimal resonant peaks from the spectra and fit them with a Lorentz peak function, yielding Q factors of 2.1$\times$10$^3$ and 5.25$\times$10$^3$, respectively. The Q factors of the smaller microdisks with the radius of 0.75 $\mu$m and 1 $\mu$m are measured to 1.2$\times$10$^3$ and 1.6$\times$10$^3$, the decrease in Q-value for smaller microdisks can be ascribed to increased scattering losses, for its modes typically closer to the rough sidewall. The radiative Q of the 0.75$\sim$2 $\mu$m micro-disks is calculated to above 10$^5$, therefore, we expect that the current Q-value is limited by the scattering and absorption losses(a detail analysis of the limits of Q factor is given in Supplement 2), the latter possibly originate from the defects introduced by ion cut process\cite{zheng20194h}. Remarkably, the highest Q-value observed in the microdisk resonators is comparable to those observed in wavelength-sized microdisk fabricated by 3C-SiC material\cite{radulaski2015visible,allioux2018toward}.\\
\indent~Another key figure of merit of the wavelength-sized microdisk resonators is the $Q/V$ ratio\cite{purcell1995spontaneous}. $V$ is the cavity mode volume, and it is defined by $V =\int[\epsilon(r)E(r)^2]d^3r/max[\epsilon(r) E(r)^2]$, with $\epsilon(r)$ being the dielectric constant, $E(r)$ the electric field intensity. Considering the highest Q-value observed at the wavelength of 862.1 nm, a mode volume of 2.61$\times (\lambda/n)^3$ is extracted from the finite-element-method (FEM) simulation. Given the Q-value of 5.25$\times$10$^3$, the $Q/V$ ratio is found to comparable with the value observed in the PhC nanobeam cavity in the bulk crystalline 4H-SiC\cite{crook2020purcell}. With these results presented above, we emphasize that our work has, for the first time, realized microdisk cavities which promise to interface the related visible and infrared spin defects (i.e., silicon related vacancy centers) in 4H-SiCOI material.
\begin{figure}[htbp]
\centering
\fbox{\includegraphics{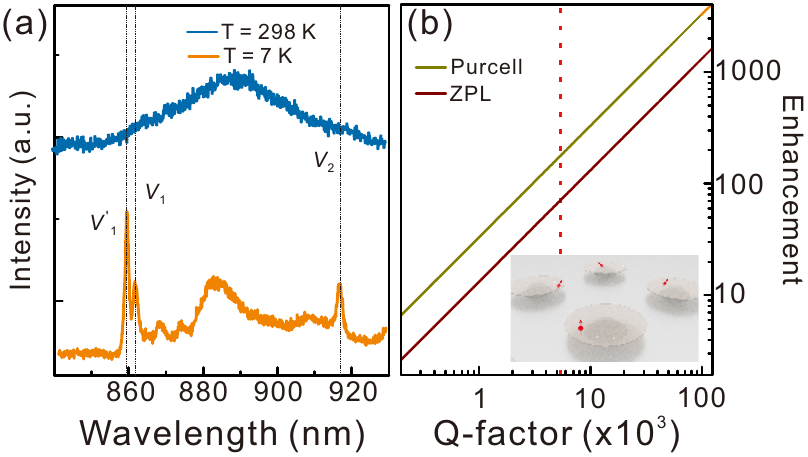}}
\caption{(a) Temperature dependent spectra from the 4H-SiCOI sample. (b) Estimated Purcell enhancement and $V_1$ ZPL emission enhancement factor as a function of optical Q factor for a 4H-SiC microdisk. The dashed line shows our current device value.}
\label{fig4}
\end{figure}
\indent~So far, we have realized WGMs in our microdisk resonators working at the visible and near infra-red wavelengths. The successful fabrication of these devices is strictly connected to the capability of achieving an effective defect-cavity system on 4H-SiCOI, a nontrivial task that is discussed in the following. It has been well known that 4H-SiC hosts a number of spin defects with the ZPL in the visible and near infrared spectral range. Figure \ref{fig4} shows temperature dependent spectra of our 4H-SiCOI sample from which distinct emission of silicon vacancy center at 858.7 nm($V_1'$) , 862 nm ($V_1$) and 917 nm ($V_2$) is observed. When a precise overlap between the WGMs and the ZPL line is established, the Purcell enhancement to the ZPL emission can be expected and the enhancement factor is evaluated by the Weisskopf-Wigner spontaneous emission theory\cite{santori2010nanophotonics,walls2007quantum}:
\begin{equation}
F_{se} = \frac{4g^2}{\kappa\gamma_{tatol}} = \frac{3}{4\pi^2}\frac{Q}{V}(\frac{\lambda}{n})^3\frac{E(r_{defect})}{E(r_{max})}\frac{\gamma_{ZPL}}{\gamma_{total}},
\label{eq1}
\end{equation}
\noindent where \emph{g} represents the photon coupling strength, $\kappa$ is the decay rate of the cavity, $\gamma_{total}$ is the total emission rate of the defect, $\gamma_{ZPL}$ is the ZPL emission rate, $E(r_{max})$ represents the maximum strength field of the cavity mode, $E(r_{defect})$ is the electric field at the defect site. For $V_1$ in 4H-SiC, the ZPL emission rate is about 40$\%$ of total emission rate at temperature of 4K\cite{nagy2018quantum}. Provided that a unity overlaps of the emitter’s dipole with the cavity mode $E(r_{defect})/ E(r_{max})$ = 1, the emission enhancement factor $F_{SE}$ of about 60 can be achieved within the current microdisk. For further enhancement of the ZPL emission, higher Q-value microcavities are needed. Figure \ref{fig4}(b) plot the enhancement factor of the $V_1$ emission as a function of Q-value, and the dashed line represents the value for the current device. It is shown that when the Q-value reaches 10$^5$, the Purcell factor can be improved up to 4000. In this scenario, the ZPL emission will be enhanced by 3 orders of magnitude. This remarkable enhancement could pave the way towards the realization of an efficient defect-cavity system where single-photon emission as well as the optically-addressable spin state readout can be achieved. 
\section{\label{sec:level3} Conclusion and outlooks}
\indent~In conclusion, we have realized wavelength-sized microdisk resonators on 4H-SiCOI formed by the ion cut process. The reported cavities support WGMs with high quality factors up to 5.25$\times$10$^3$ and small mode volume of about 2.61$\times(\lambda/n)^3$ at the visible and near-infrared wavelengths, and its  resonances compatible with the ZPL lines of spin-defects in 4H-SiCOI. Therefore, the wavelength-sized microdisk resonators demonstrated in our work can be envisaged to develop an effective defect-cavity system for potential applications in quantum information processing and quantum sensing.\\
\indent~Nevertheless, work is needed to further improve the device to achieve the maximum capability in the context of quantum applications. This includes preparation of single silicon vacancy center and external control of the cavity resonance modes. One of the possible strategies to prepare single silicon vacancy center is to elevate the annealing temperature for recovering proton-irradiated defects so as to further reduce the density of color center. Note that this strategy is already proved to be efficient for improving the crystalline quality of the 4H-SiCOI after the proton implantation\cite{yi2020wafer}. As for spectral tuning of the cavity , thermal heating or annealing methods\cite{wu2019high} provides a primary choice. Aside from this, numerous other techniques such as cryogenic gas condensation\cite{lukin20204h,crook2020purcell}, atomic layer deposition and laser-assisted oxidation\cite{kiravittaya2011tuning} can in principle be applied to our micro-disk resonator. 

\nocite{*}
\bibliography{manuscript}
\end{document}